\documentclass{article}%
\usepackage{amsmath}
\usepackage{amssymb}
\usepackage{amsfonts}
\usepackage{graphicx}%
\providecommand{\U}[1]{\protect\rule{.1in}{.1in}}

\begin{document}

\title{The Relation between Maxwell, Dirac and the Seiberg-Witten
Equations\thanks{published: \textit{International Journal of Mathematics and
Mathematical Sciences} \textbf{2003}, 2007-2034 (2003). This version corrects
several misprints and some typos.}}
\author{Waldyr A. Rodrigues, Jr.\\\hspace{-0.8cm} {\small Institute of Mathematics, Statitistics and Scientific
Computation}\\{\small \ IMECC-UNICAMP CP 6065}\\{\small \ 13083-970 Campinas SP Brazil}\\{\small \ e-mail: walrod@ime.unicamp.br}\\{\small key words: Maxwell equations, Dirac equation, Seiberg-Witten
equation.}}
\date{received 10/22/2002, revised 12/26/2002}
\maketitle

\begin{abstract}
In this paper we discuss some unusual and unsuspected relations between
Maxwell, Dirac and the Seiberg-Witten equations. First we investigate what is
now known as the Maxwell-Dirac equivalence (\emph{MDE}) of the first kind.
Crucial to that proposed equivalence is the possibility of solving for $\psi$
(a representative on a given spinorial frame of a Dirac-Hestenes spinor field)
the equation $F=\psi\gamma_{21}\tilde{\psi}$, where $F$ is a given
electromagnetic field. Such non trivial task is presented in this paper and it
permits to clarify some possible objections to the \emph{MDE} which claims
that no \emph{MDE} may exist, because $F$ has six (real) degrees of freedom
and $\psi$ has eight (real) degrees of freedom. Also, we review the
generalized Maxwell equation describing charges and monopoles. The enterprise
is worth even if there is \emph{no} evidence until now for magnetic monopoles,
because there are at least two faithful field equations that have the form of
the generalized Maxwell equations. One is the generalized Hertz potential
field equation (which we discuss in detail) associated with Maxwell theory and
the other is a (non linear) equation (of the generalized Maxwell type)
satisfied by the 2-form field part of a Dirac-Hestenes spinor field that
solves the Dirac-Hestenes equation for a free electron. This is a new and
surprising result, which can also be called \emph{MDE} of the second kind. It
strongly suggests that the electron is a composed system with more elementary
``charges'' of the electric and magnetic types. This finding may eventually
account for the recent claims that the electron has been splited into two
electrinos. Finally, we use the \emph{MDE} of the first kind together with a
reasonable hypothesis to give a derivation of the famous Seiberg-Witten
equations on Minkowski spacetime. A suggestive physical interpretation for
those equations is also given.

\end{abstract}
\tableofcontents

\section{Introduction}

In (\cite{1}-\cite{5}) using standard covariant spinor fields Campolattaro
proposed that Maxwell equations are equivalent to a \emph{non} linear Dirac
like equation. The subject has been further developed in (\cite{6},\cite{8})
using the Clifford bundle formalism, which is discussed together with some of
their applications in a series of papers, e.g., (\cite{6}-\cite{19}). The
crucial point in proving the mentioned equivalence (abbreviated as \emph{MDE}
\ in what follows, when no confusion arises), starts once we observe that to
any given \emph{representative} of a Dirac-Hestenes spinor field (see more
information see section 2 and for details see (\cite{12},\cite{14}%
,\cite{16},\cite{17}) $\psi\in\sec[\bigwedge^{0}(M)+\bigwedge^{2}%
(M)+\bigwedge^{4}(M)]\subset\sec\mathcal{C}\!\ell(M,g)$ there is associated an
electromagnetic field $F\in\sec\bigwedge^{2}(M)\subset\sec\mathcal{C}%
\!\ell(M,g)$, ($F^{2}\neq0$) through the Rainich-Misner theorem (\cite{20}%
,\cite{6}-\cite{8}) by
\begin{equation}
F=\psi\gamma_{21}\tilde{\psi} \label{1}%
\end{equation}

Before proceeding we recall that for null fields, i.e., $F^{2}=0$, the spinor
associated with $F$ through Eq.(\ref{1}) must be a Majorana spinor field
(\cite{6},\cite{14},\cite{15}), but we do not need such concept in this paper.
Now, since an electromagnetic field $F$ satisfying Maxwell \textit{equation}
has six degrees of freedom and a Dirac-Hestenes spinor field has eight (real)
degrees of freedom some authors felt uncomfortable with the approach used in
(\cite{7},\cite{8}) where some \emph{gauge conditions} have been imposed on a
\emph{nonlinear} equation (equivalent to Maxwell equation), thereby
transforming it into an usual \emph{linear} Dirac equation (called the
\emph{Dirac-Hestenes equation} in the Clifford bundle formalism).The claim,
e.g., in \cite{21} is that the \emph{MDE} found in (\cite{7}, \cite{8}) cannot
be general. The argument is that the imposition of \emph{gauge conditions}
implies that a $\psi$ satisfying Eq.(\ref{1}) can have only six (real) degrees
of freedom, and this implies that the Dirac-Hestenes equation corresponding to
Maxwell equation can be only satisfied by a restricted class of Dirac-Hestenes
spinor fields, namely the ones that have six degrees of freedom.

Incidentally, in \cite{21} it is also claimed that the generalized Maxwell
equation
\begin{equation}
{\mbox{\boldmath$\partial$}}F=J_{e}+\gamma_{5}J_{m} \label{2}%
\end{equation}
(where $J_{e},$ $J_{m}\in\sec\bigwedge^{1}(M)$) describing the electromagnetic
field generated by charges and monopoles \cite{9} cannot hold in the Clifford
bundle formalism, because according to that author the formalism implies that
$J_{m}=0$.

In what follows we analyze these claims of \cite{21} and prove that they are
wrong (section 3). The reasons for our enterprise is that as will become clear
in what follows, understanding of Eqs.(\ref{1}) and (\ref{2}) together with
some reasonable hypothesis permit a derivation and even a possible physical
interpretation of the famous Seiberg-Witten monopole equations (\cite{22}%
,\cite{23},\cite{26}). So, our plan is the following: first we introduce in
section 2 the mathematical formalism used in the paper, showing how to write
Maxwell and Dirac equations using Clifford fields. We also introduce Weyl
spinor fields and parity operators in the Clifford bundle formalism. In
section 3 we prove that given $F$ in Eq.(\ref{1}) we can solve that equation
for $\psi$, and we find that $\psi$ has eight degrees of freedom, two of them
being undetermined, the indetermination being related to the elements of the
\textit{stability group} of the spin plane $\gamma_{21}$. This is a non
trivial and beautiful result which can called \emph{inversion} formula. In
section 4 we introduce a \emph{generalized} Maxwell equation and in section 5
we introduce the \emph{generalized} Hertz equation. In section 6 we prove a
\textit{mathematical} Dirac-Maxwell equivalence of the \emph{first kind}
(\cite{1},\cite{8}), thereby deriving a Dirac-Hestenes equation from the free
Maxwell equations. In section 7 we introduce a new form of a
\textit{mathematical} Maxwell-Dirac equivalence (called \emph{MDE }of the
second kind) different from the one studied in section 6. This new \emph{MDE}
of the second kind suggests that the electron is a \ `composite' system. To
prove the Maxwell-Dirac equivalence of the \emph{second kind} we decompose a
Dirac-Hestenes spinor field satisfying a Dirac-Hestenes equation in such a way
that it results in a nonlinear generalized Maxwell (like) equation
(Eq.(\ref{7.1})) satisfied by a certain Hertz potential field, mathematically
represented by an object of the same mathematical nature as an electromagnetic
field, i.e., $\Pi\in\sec\bigwedge^{2}(M)\subset\sec\mathcal{C}\!\ell
(M,g)$.This new equivalence is very suggestive in view of the fact that there
are recent (wild) speculations that the electron can be splited in two
components \cite{27} (see also\cite{28}). If this fantastic claim announced by
Maris \cite{27} is true, it is necessary to understand what is going on. The
new Maxwell-Dirac equivalence presented in section 6 may eventually be useful
to understand the mechanism behind the \textquotedblleft electron
splitting\textquotedblright\ into electrinos. We are not going to discuss
these ideas here. Instead, we concentrate our attention in showing in section
8 that (the \textit{analogous} on Minkowski spacetime) of the famous
Seiberg-Witten monopole equations arises naturally from the \emph{MDE} of the
first kind once a reasonable hypothesis is imposed. We also present a possible
coherent interpretation of that equations. Indeed, we prove that when the
Dirac-Hestenes spinor field satisfying the first of Seiberg-Witten equations
is an eigenvector of the parity operator, then that equation describe a pair
of massless `monopoles' of opposite `magnetic' like charges, coupled together
by its interaction electromagnetic field. Finally, in section 9 we present our conclusions.

\section{Clifford and Spin-Clifford Bundles}

Let $\mathcal{M}=(M,g,D)$ be Minkowski spacetime. $(M,g)$ is a four
dimensional time oriented and space oriented Lorentzian manifold, with
$M\simeq\mathbb{R}^{4}$ and $g\in\sec T^{0,2}M$ being a Lorentzian metric of
signature (1,3). $T^{\ast}M$ [$TM$] is the cotangent [tangent] bundle.
$T^{\ast}M=\cup_{x\in M}T_{x}^{\ast}M$, $TM=\cup_{x\in M}T_{x}M$, and
$T_{x}M\simeq T_{x}^{\ast}M\simeq\mathbb{R}^{1,3}$, where $\mathbb{R}^{1,3}$
is the Minkowski vector space~. $D$ is the Levi-Civita connection of $g$,
$i.e$\textit{.\/}, $Dg=0$, $\mathbf{R}(D)=0$. Also $\mathbf{T}(D)=0$,
$\mathbf{R}$ and $\mathbf{T}$ being respectively the torsion and curvature
tensors. Now, the Clifford bundle of differential forms $\mathcal{C}%
\!\ell(M,g)$ is the bundle of algebras, i.e., $\mathcal{C}\!\ell
(M,g)=\cup_{x\in M}\mathcal{C}\!\ell(T_{x}^{\ast}M)$, where $\forall x\in
M,\mathcal{C}\!\ell(T_{x}^{\ast}M)=\mathcal{C}\!\ell_{1,3}$, the so called
\emph{spacetime} \emph{algebra}. Recall also that $\mathcal{C}\!\ell(M,g)$ is
a vector bundle associated to the \emph{\ orthonormal frame bundle}, i.e.,
$\mathcal{C}\!\ell(M,g)$ $=P_{SO_{+(1,3)}}\times_{ad}\mathcal{C}l_{1,3}$
(\cite{16},\cite{17}). For any $x\in M$, $\mathcal{C}\!\ell(T_{x}^{\ast}M)$ as
a linear space over the real field $\mathbb{R}$. Moreover, $\mathcal{C}%
\!\ell(T_{x}^{\ast}M)$ is isomorphic to the Cartan algebra $\bigwedge
(T_{x}^{\ast}M)$ of the cotangent space and $\bigwedge(T_{x}^{\ast}%
M)=\sum_{k=0}^{4}\bigwedge{}^{k}(T_{x}^{\ast}M)$, where $\bigwedge^{k}%
(T_{x}^{\ast}M)$ is the $\binom{4}{k}$-dimensional space of $k$-forms. Then,
sections of $\mathcal{C}\!\ell(M,g)$ can be represented as a sum of non
homogeneous differential forms. Let $\langle x^{\mu}\rangle$ be Lorentz
coordinate functions for $M$ and let $\{e_{\mu}\}\in\sec FM$ (the frame
bundle) be an orthonormal basis for $TM$, i.e., $g(e_{\mu},e_{\nu})=\eta
_{\mu\nu}=\mathrm{diag}(1,-1,-1,-1)$. Let $\gamma^{\nu}=dx^{\nu}\in
\sec\bigwedge^{1}(M)\subset\sec\mathcal{C}\!\ell(M,g)$ ($\nu=0,1,2,3$) such
that the set $\{\gamma^{\nu}\}$ is the dual basis of $\{e_{\mu}\}$. Moreover,
we denote by $\check{g}$ the metric in the cotangent bundle.

\subsection{Clifford Product}

The fundamental \emph{Clifford product} (in what follows to be denoted by
juxtaposition of symbols) is generated by $\gamma^{\mu}\gamma^{\nu}%
+\gamma^{\nu}\gamma^{\mu}=2\eta^{\mu\nu}$ and if $\mathcal{C}\in
\sec\mathcal{C}\!\ell(M,g)$ we have%

\begin{equation}
\mathcal{C}=s+v_{\mu}\gamma^{\mu}+\frac{1}{2!}b_{\mu\nu}\gamma^{\mu}%
\gamma^{\nu}+\frac{1}{3!}a_{\mu\nu\rho}\gamma^{\mu}\gamma^{\nu}\gamma^{\rho
}+p\gamma^{5}\;, \label{3}%
\end{equation}
where $\gamma^{5}=\gamma^{0}\gamma^{1}\gamma^{2}\gamma^{3}=dx^{0}dx^{1}%
dx^{2}dx^{3}$ is the volume element and $s$, $v_{\mu}$, $b_{\mu v}$,
$a_{\mu\nu\rho}$, $p\in\sec\bigwedge^{0}(M)\subset\sec\mathcal{C}\!\ell(M,g)$.

Let $A_{r},\in\sec\bigwedge^{r}(M),B_{s}\in\sec\bigwedge^{s}(M)$. For $r=s=1$,
we define the \emph{scalar product} as follows:

For $a,b\in\sec\bigwedge^{1}(M)\subset\sec\mathcal{C}\!\ell(M,g).,$%
\begin{equation}
a\cdot b=\frac{1}{2}(ab+ba)=\check{g}(a,b). \label{4}%
\end{equation}
We define also the \emph{exterior product} ($\forall r,s=0,1,2,3)$ by
\begin{align}
A_{r}\wedge B_{s}  &  =\langle A_{r}B_{s}\rangle_{r+s},\nonumber\\
A_{r}\wedge B_{s}  &  =(-1)^{rs}B_{s}\wedge A_{r}, \label{5}%
\end{align}
where $\langle\rangle_{k}$ is the component in $\bigwedge^{k}(M)$ of the
Clifford field. The exterior product is extended by linearity to all sections
of $\mathcal{C}\!\ell(M,g)$.

For $A_{r}=a_{1}\wedge...\wedge a_{r},B_{r}=b_{1}\wedge...\wedge b_{r}$, the
scalar product is defined here as follows,
\begin{align}
A_{r}\cdot B_{r}  &  =(a_{1}\wedge...\wedge a_{r})\cdot(b_{1}\wedge...\wedge
b_{r})\nonumber\\
&  =\left\vert
\begin{array}
[c]{lll}%
a_{1}\cdot b_{1} & .... & a_{1}\cdot b_{r}\\
.......... & .... & ..........\\
a_{r}\cdot b_{1} & .... & a_{r}\cdot b_{r}%
\end{array}
\right\vert \label{6}%
\end{align}

We agree that if $r=s=0$, the scalar product is simple the ordinary product in
the real field.

Also, if $r\neq s$, then $A_{r}\cdot B_{s}=0$. Finally, the scalar product is
extended by linearity for all sections of $\mathcal{C}\!\ell(M,g)$.

For $r\leq s,A_{r}=a_{1}\wedge...\wedge a_{r},B_{s}=b_{1}\wedge...\wedge
b_{s\text{ }}$we define the left contraction by
\begin{equation}
\lrcorner:(A_{r},B_{s})\mapsto A_{r}\lrcorner B_{s}=%
{\displaystyle\sum\limits_{i_{1}\,<...\,<i_{r}}}
\epsilon^{i_{1}....i_{s}}(a_{1}\wedge...\wedge a_{r})\cdot(b_{_{i_{1}}}%
\wedge...\wedge b_{i_{r}})^{\sim}b_{i_{r}+1}\wedge...\wedge b_{i_{s}}
\label{7}%
\end{equation}
where $\sim$ is the reverse mapping (\emph{reversion}) defined by
\begin{equation}
\sim:\sec\bigwedge^{p}(M)\ni a_{1}\wedge...\wedge a_{p}\mapsto a_{p}%
\wedge...\wedge a_{1} \label{8}%
\end{equation}
and extended by linearity to all sections of $\mathcal{C}\!\ell(M,g)$. We
agree that for $\alpha,\beta\in\sec\bigwedge^{0}(M)$ the contraction is the
ordinary (pointwise) product in the real field and that if $\alpha\in
\sec\bigwedge^{0}(M)$, $A_{r},\in\sec\bigwedge^{r}(M),B_{s}\in\sec
\bigwedge^{s}(M)$ then $(\alpha A_{r})\lrcorner B_{s}=A_{r}\lrcorner(\alpha
B_{s})$. Left contraction is extended by linearity to all pairs of elements of
sections of $\mathcal{C}\!\ell(M,g)$, i.e., for $A,B\in\sec\mathcal{C}%
\!\ell(M,g)$%

\begin{equation}
A\lrcorner B=\sum_{r,s}\langle A\rangle_{r}\lrcorner\langle B\rangle_{s},r\leq
s \label{9}%
\end{equation}

It is also necessary to introduce the operator of \emph{right contraction}
denoted by $\llcorner$. The definition is obtained from the one presenting the
left contraction with the imposition that $r\geq s$ and taking into account
that now if $A_{r},\in\sec\bigwedge^{r}(M),B_{s}\in\sec\bigwedge^{s}(M)$ then
$A_{r}\llcorner(\alpha B_{s})=(\alpha A_{r})\llcorner B_{s}$.

The main formulas used in the Clifford calculus can be obtained from the
following ones (where $a\in\sec\bigwedge^{1}(M)\subset\sec\mathcal{C}%
\!\ell(M,g)$):
\begin{align}
aB_{s}  &  =a\lrcorner B_{s}+a\wedge B_{s},B_{s}a=B_{s}\llcorner a+B_{s}\wedge
a,\nonumber\\
a\lrcorner B_{s}  &  =\frac{1}{2}(aB_{s}-(-)^{s}B_{s}a),\nonumber\\
A_{r}\lrcorner B_{s}  &  =(-)^{r(s-1)}B_{s}\llcorner A_{r},\nonumber\\
a\wedge B_{s}  &  =\frac{1}{2}(aB_{s}+(-)^{s}B_{s}a),\nonumber\\
A_{r}B_{s}  &  =\langle A_{r}B_{s}\rangle_{|r-s|}+\langle A_{r}\lrcorner
B_{s}\rangle_{|r-s-2|}+...+\langle A_{r}B_{s}\rangle_{|r+s|}\nonumber\\
&  =\sum\limits_{k=0}^{m}\langle A_{r}B_{s}\rangle_{|r-s|+2k}\nonumber\\
A_{r}\cdot B_{r}  &  =B_{r}\cdot A_{r}=\tilde{A}_{r}\lrcorner B_{r}%
=A_{r}\llcorner\tilde{B}_{r}=\langle\tilde{A}_{r}B_{r}\rangle_{0}=\langle
A_{r}\tilde{B}_{r}\rangle_{0} \label{10}%
\end{align}

\subsubsection{Hodge Star Operator}

Let $\star$ be the Hodge star operator, i.e., the mapping
\[
\star:\bigwedge^{k}(M)\rightarrow\bigwedge^{4-k}(M),\text{ }A_{k}\mapsto\star
A_{k}%
\]
where for $A_{k}\in\sec\bigwedge^{k}(M)\subset\sec\mathcal{C}\!\ell(M,g)$%
\begin{equation}
\lbrack B_{k}\cdot A_{k}]\tau_{g}=B_{k}\wedge\star A_{k},\forall B_{k}\in
\sec\bigwedge\nolimits^{k}(M)\subset\sec\mathcal{C}\!\ell(M,g). \label{11a}%
\end{equation}
$\tau_{g}\in\bigwedge^{4}(M)$ is a \emph{standard} volume element. Then we can
verify that
\begin{equation}
\star A_{k}=\widetilde{A}_{k}\gamma^{5}. \label{11b}%
\end{equation}

\subsubsection{Dirac Operator}

Let $d$ and $\delta$ be respectively the differential and Hodge codifferential
operators acting on sections of $\bigwedge(M)$. If $A_{p}\in\sec\bigwedge
^{p}(M)\subset\sec\mathcal{C}\!\ell(M,g)$, then $\delta A_{p}=(-)^{p}%
\star^{-1}d\star A_{p}$, with $\star^{-1}\star=\mathrm{identity}$.

The Dirac operator acting on sections of $\mathcal{C}\!\ell(M,g)$ is the
invariant first order differential operator
\begin{equation}
{\mbox{\boldmath$\partial$}}=\gamma^{a}D_{e_{a}}, \label{12}%
\end{equation}
where $\{e_{a}\}$ is an arbitrary \emph{orthonormal basis} for $TU\subset TM$
and $\{\gamma_{b}\}$ is a basis for $T^{\ast}U\subset T^{\ast}M$ dual to the
basis $\{e_{a}\}$, i.e., $\gamma^{b}(e_{a})=\delta_{b}^{a}$, $a,b=0,1,2,3$.
The reciprocal basis of $\{\gamma^{b}\}$ is denoted $\{\gamma_{a}\}$ and we
have $\gamma_{a}\cdot\gamma_{b}=\eta_{ab}$ ($\eta_{ab}=\mathrm{diag}%
(1,-1,-1,-1)$). Also,
\begin{equation}
D_{e_{a}}\gamma^{b}=-\omega_{a}^{bc}\gamma_{c} \label{12n}%
\end{equation}
Defining
\begin{equation}
\mathbf{\omega}_{a}=\frac{1}{2}\omega_{a}^{bc}\gamma_{b}\wedge\gamma_{c},
\label{12nn}%
\end{equation}
we have that for any $A_{p}\in\sec\bigwedge^{p}(M),$ $p=0,1,2,3,4$
\begin{equation}
D_{e_{a}}A=e_{a}+\frac{1}{2}[\mathbf{\omega}_{a},A]. \label{12nnn}%
\end{equation}

Using Eq.(\ref{12nnn}) we can show the very important result:%

\begin{align}
{\mbox{\boldmath$\partial$}}A_{p}  &  ={\mbox{\boldmath$\partial$}}\wedge
A_{p\,}+\,{\mbox{\boldmath$\partial$}}\lrcorner A_{p}=dA_{p}-\delta
A_{p},\nonumber\\
{\mbox{\boldmath$\partial$}}\wedge A_{p}  &  =dA_{p},\hspace{0.1in}%
\,{\mbox{\boldmath$\partial$}}\lrcorner A_{p}=-\delta A_{p}, \label{13}%
\end{align}

\subsection{Dirac-Hestenes Spinor Fields}

Now, as is well known, an electromagnetic field is represented by $F\in
\sec\bigwedge^{2}(M)\subset\sec\mathcal{C}\!\ell(M,g)$. How to represent the
Dirac spinor fields in this formalism~? We can show that \emph{Dirac-Hestenes}
spinor fields, do the job. \ We give here a short introduction to these
objects (when living on Minkowski spacetime) which serves mainly the purpose
of fixing notations. For a rigorous theory of these objects (using vector
bundles) on a general Riemann-Cartan manifold see (\cite{17}). Recall that
there is a $2:1$ mapping $\mathbf{s}^{\prime}:\mathbf{\Theta}^{\prime
}\mathbf{\rightarrow}\mathcal{B}$ between $\mathcal{B}$, the set of all
orthonormal ordered vector frames and $\mathbf{\Theta}^{\prime}$, the set of
all \textit{spin frames} of $T^{\ast}M$. As discussed at length in
(\cite{16},\cite{17}) a spin \textit{coframe} can be thought as a basis of
$T^{\ast}M$, such that two ordered basis \ even if consisting of the same
vectors, but, with the spatial vectors differing by a $2\pi$ rotation are
considered \textit{distinct} and two ordered basis even if consisting of the
same vectors, but \ with the spatial vectors differing by a $4\pi$ rotation
are identified. For short, in this paper we call the spin coframes, simply
\textit{spin frames}. Also, vector coframes are simply called vector frames in
what follows.

Consider the set $\mathcal{S}$ of mappings
\begin{equation}
M\ni x\mapsto u(x)\in\mathrm{Spin}_{+}(1,3)\text{ }\mathbb{\simeq}\text{
}Sl(2,\mathbb{C}) \label{13b}%
\end{equation}
Choose a constant spin frame $\{\mathbf{\gamma}_{a}\}\in\mathcal{B}$,
$a=0,1,2,3$ and choose $\Xi_{0}\in\mathbf{\Theta}^{\prime}$ corresponding to a
constant mapping $u_{0}\in\mathcal{S}$. By constant we mean that the equation
$\mathbf{\gamma}_{\mu}(x)=\mathbf{\gamma}_{\mu}(y)$ ($(\mu=0,1,2,3)$ and
$u_{0}(x)=u_{0}(y),$ $\forall$ $x,y\in M$) has meaning due to the usual
\textit{affine} structure that can be given to Minkowski spacetime. $\Xi
_{0},\Xi_{u}\in\mathbf{\Theta}^{\prime}$ are relate as follows
\begin{equation}
u_{0}\mathbf{s}^{\prime}(\Xi_{0})u_{0}^{-1}=u\mathbf{s}^{\prime}(\Xi
_{u})u^{-1} \label{13c}%
\end{equation}

From now on in order to simplify the notation we take $u_{0}=1$. The frame
$\mathbf{s}^{\prime}(\Xi_{0})=\{\mathbf{\gamma}_{a}\}$ is called the
\emph{fiducial} \ vector frame and $\Xi_{0\text{ }}$the fiducial \textit{spin
}frame. We note that Eq.(\ref{13c}) is satisfied by \emph{two} such $u$'s
differing by a signal, and of course, $\mathbf{s}^{\prime}(\Xi_{u}%
)=\mathbf{s}^{\prime}(\Xi_{-u})$.

Let,
\begin{equation}
\mathfrak{T}=\{(\Xi_{u},\Psi_{\Xi_{u}})\text{ }|\text{ }u\in\mathcal{S}%
,\Xi_{u}\in\mathbf{\Theta}^{\prime},\Psi_{\Xi_{u}}\in\sec\bigwedge
\limits^{+}M\subset\sec C\ell^{+}(M,g)\}, \label{13d}%
\end{equation}
where $\bigwedge\limits^{+}M=\bigwedge^{0}M+\bigwedge^{2}M+\bigwedge^{4}M$

We define an equivalence relation on $\mathfrak{T}$ by setting
\begin{equation}
(\Xi_{u},\Psi_{\Xi_{u}})\sim(\Xi_{u^{\prime}},\Psi_{\Xi_{u^{\prime}}})
\label{13e}%
\end{equation}
if and only if
\begin{equation}
u\mathbf{s}^{\prime}(\Xi_{u})u^{-1}=u^{\prime-1}\mathbf{s}^{\prime}%
(\Xi_{u^{\prime}})u^{\prime},\text{ }\Psi_{\Xi_{u^{\prime}}}=\Psi_{\Xi_{u}%
}uu^{\prime-1}. \label{13f}%
\end{equation}

\textbf{Definition}: Any equivalence class $\mathbb{[(}\Xi_{u},\Psi_{\Xi_{u}%
})\mathbf{]}$ will be called a Dirac-Hestenes spinor field.

Before proceeding we recall that a more rigorous definition of a \emph{DHSF}
as a section of a spin-Clifford bundle is given in \cite{17}. We will not need
such a sophistication in what follows.

We observe that the pairs $(\mathbf{\Xi}_{u},\Psi_{\mathbf{\Xi}_{u}})$ and
$(\mathbf{\Xi}_{-u},\Psi_{\mathbf{\Xi}_{-u}})=(\mathbf{\Xi}_{-u}%
,-\Psi_{\mathbf{\Xi}_{u}})$ are equivalent, but the pairs $(\mathbf{\Xi}%
_{u},\Psi_{\mathbf{\Xi}_{u}})$ and $(\mathbf{\Xi}_{-u},\Psi_{\mathbf{\Xi}%
_{-u}})=(\mathbf{\Xi}_{-u},\Psi_{\mathbf{\Xi}_{u}})$ are not. This distinction
is essential in order to give a structure of \emph{linear space} (over the
real numbers) to the set $\mathcal{T}$. Indeed, such a linear structure on
$\mathcal{T}$ is defined as follows
\begin{align}
\hspace{0.1in}\hspace{-0.09in}a[(\Xi_{u_{1}},\Psi_{\Xi_{u_{1}}})]+b[(\Xi
_{u_{2}},\Psi_{\Xi_{u_{2}}})]\hspace{-0.1in}  &  =\hspace{-0.1in}[(\Xi_{u_{1}%
},a\Psi_{\Xi_{u_{1}}})]+[(\Xi_{u_{2}},b\Psi_{\Xi_{u_{2}}})],\nonumber\\
\hspace{-0.09in}(a+b)[(\Xi_{u_{1}},\Psi_{\Xi_{u_{1}}})]\hspace{-0.1in}  &
=\hspace{-0.1in}a[(\Xi_{u_{1}},\Psi_{\Xi_{u_{1}}})]+b[(\Xi_{u_{1}},\Psi
_{\Xi_{u_{1}}})],\nonumber\\
a,b  &  \in\mathbb{R}. \label{13g}%
\end{align}

We can simplify the notation by recalling that every $u\in\mathcal{S}$
determines, of course, a unique spin frame $\Xi_{u}$ . Taking this into
account we consider the set of all pairs $(u,\Psi_{\Xi_{u}})\in\mathcal{S}%
\mathbb{\times}\sec\mathcal{C\ell}^{+}(M,g)$

We define an equivalence relation $\mathcal{R}$ in $\ \mathcal{S}%
\mathbb{\times}\sec\mathcal{C\ell}^{+}(M,g)$ as follows. Two pairs
$(u,\Psi_{\Xi_{u}})$, $(u^{\prime},\Psi_{\Xi_{u^{\prime}}})\in\sec
\mathcal{S}\mathbb{\times}\sec\mathcal{C\ell}^{+}(M,g)$ are equivalent if and
only if
\begin{equation}
\Psi_{\Xi_{u^{\prime}}}u^{\prime}=\Psi_{\Xi_{u}}u \label{13h}%
\end{equation}

Of course, $\mathbf{s}^{\prime}(\Xi_{u^{\prime}})=v\mathbf{s}^{\prime}(\Xi
_{u})v^{-1}$with $v=(u^{\prime})^{-1}u\in\mathcal{S}$. Note that the pairs
$(u,\Psi_{\Xi_{u}})$ and $(-u,-\Psi_{\Xi_{u}})$ are equivalent but the pairs
$(u,\Psi_{\Xi_{u}})$ and $(-u,\Psi_{\Xi_{u}})$ are not.

Denote by $\mathcal{S}\mathbb{\times}\sec\mathcal{C\ell}^{+}(M,g)$
$/\mathcal{R}$ the quotient set of the equivalence classes generated by
$\mathcal{R}$. Their elements are called Dirac-Hestenes\emph{\ spinors}. Of
course, this is the same definition as above.

From now on we simplify even more our notation. In that way, if we take two
orthonormal spin frames $\mathbf{s}^{\prime}(\Xi)=\{\gamma^{\mu}\}$ and
$\mathbf{s}^{\prime}\mathbf{(}\dot{\Xi})=\{\dot{\gamma}^{\mu}=R\gamma^{\mu
}\widetilde{R}=\Lambda_{\nu}^{\mu}\gamma^{\nu}\}$ with $\Lambda_{\nu}^{\mu
}(x)\in\mathrm{SO}^{e}(1,3)$ and $R(x)\in\mathrm{Spin}^{e}(1,3)\;\forall x\in
M$, $R\widetilde{R}=\widetilde{R}R=1$, then we simply write the relation
(Eq.(\ref{13h})) between representatives of a Dirac-Hestenes spinor field in
the two spin frames as the sections $\psi_{\Xi}$ and $\psi_{\dot{\Xi}}$ of
$\mathcal{C}\!\ell^{+}(M,g)$ related by
\begin{equation}
\psi_{\dot{\Xi}}=\psi_{\Xi}R. \label{14}%
\end{equation}
Recall that since $\psi_{\Xi}\in\sec\bigwedge\limits^{+}M\subset
\sec\mathcal{C\ell}^{+}(M,g)$, we have
\begin{equation}
\psi_{\Xi}=s+\frac{1}{2!}b_{\mu\nu}\gamma^{\mu}\gamma^{\nu}+p\gamma^{5}.
\label{15}%
\end{equation}
Note that $\psi_{\Xi}$ has the correct number of degrees of freedom in order
to represent a \emph{Dirac} spinor field and recall that the specification of
$\psi_{\Xi}$ depends on the spin frame $\Xi$. To simplify even more our
notation, when it is clear which is the spin frame $\Xi$, and no possibility
of \emph{confusion} arises we write simply $\psi$ instead of $\psi_{\Xi}$.

When $\psi\tilde{\psi}\neq0$, where $\sim$ is the reversion operator, we can
show that $\psi$ has the following canonical decomposition:
\begin{equation}
\psi=\sqrt{\rho}\,e^{\beta\gamma_{5}/2}R\,, \label{16}%
\end{equation}
where $\rho$, $\beta\in\sec\bigwedge^{0}(M)\subset\sec\mathcal{C}\!\ell(M,g)$
and $R(x)\in\mathrm{Spin}^{e}(1,3)\subset\mathcal{C}\!\ell_{1,3}^{+}$,
$\forall x\in M$. $\beta$ is called the Takabayasi angle. If we want to work
in terms of the usual Dirac spinor field formalism, we can translate our
results by choosing, for example, the standard matrix representation of the
one forms $\{\gamma^{\mu}\}$ in $\mathbb{C}(4)$ (the algebra of the complex
$4\times4$ matrices), and for $\psi_{\Sigma}$ given by Eq.(15) we have the
following (standard) matrix representation \cite{12},\cite{16}):%

\begin{equation}
\Psi=\left(
\begin{array}
[c]{cccc}%
\psi_{1} & -\psi_{2}^{*} & \psi_{3} & \psi_{4}^{*}\\
\psi_{2} & \psi_{1}^{*} & \psi_{4} & -\psi_{3}^{*}\\
\psi_{3} & \psi_{4}^{*} & \psi_{1} & -\psi_{2}^{\star}\\
\psi_{4} & -\psi_{3}^{*} & \psi_{2} & \psi_{1}^{\star}%
\end{array}
\right)  . \label{17}%
\end{equation}
where $\psi_{k}(x)\in\mathbb{C}$, $k=1,2,3,4$ and for all $x\in M$.

We recall that a \emph{standard} Dirac spinor field is a section
$\mathbf{\Psi}_{D}$ of the vector bundle $P_{\mathrm{Spin}^{e}(1,3)}%
\times_{\lambda}\mathbb{C}(4)$, where $\lambda$ is the $D(\frac{1}{2},0)\oplus
D(0,\frac{1}{2})$ representation of $Sl(2,\mathbb{C})\sim\mathrm{Spin}%
^{e}(1,3)$. For details see, e.g.,(\cite{16},\cite{17}). The relation between
$\mathbf{\Psi}_{D}$ and $\psi$ is given by
\begin{equation}
\mathbf{\Psi}_{D}=\left(
\begin{array}
[c]{c}%
\psi_{1}\\
\psi_{2}\\
\psi_{3}\\
\psi_{4}%
\end{array}
\right)  =\left(
\begin{array}
[c]{c}%
s-ib_{12}\\
-b_{13}-ib_{23}\\
-b_{03}+ip\\
-b_{01}-ib_{02}%
\end{array}
\right)  . \label{18}%
\end{equation}
where $s,\,b_{12},\ldots$ \ are the real functions in Eq.(\ref{15}) and $\ast$
denotes the complex conjugation.

We recall that the even subbundle $\mathcal{C}\!\ell^{+}(M,g)$ of
$\mathcal{C}\!\ell(M,g)$ is such that its typical fiber is the Pauli algebra
$\mathcal{C}\!\ell_{3,0}\equiv\mathcal{C}\!\ell_{1,3}^{+}$ (which is
isomorphic to $\;\mathbb{C}(2)$, the algebra of $2\times2$ complex matrices).
Elements of $\mathcal{C}\!\ell_{1,3}^{+}$ are called \emph{biquaternions} in
the old literature. The isomorphism $\mathcal{C}\!\ell_{3,0}\equiv
\mathcal{C}\!\ell_{1,3}^{+}$ is exhibited by putting $\vec{\sigma}_{i}%
=\gamma_{i}\gamma_{0}$, whence $\vec{\sigma}_{i}\vec{\sigma}_{j}+\vec{\sigma
}_{j}\vec{\sigma}_{i}=2\delta_{ij}$. We recall also that the Dirac algebra is
$\mathcal{C}\!\ell_{4,1}\equiv\;\mathbb{C(}4)$ and $\mathcal{C}\!\ell
_{4,1}\equiv\;\mathbb{C}\otimes\mathcal{C}\!\ell_{1,3}$.

Consider the \emph{complexification}\textit{\/} $\mathcal{C}\ell_{C}(M,g)$ of
$\mathcal{C}\ell(g)$ called the \emph{complex Clifford bundle}\textit{\/}.
Then $\mathcal{C}\ell_{C}(M,g)=\;\mathbb{C}\otimes\mathcal{C}\ell(M,g)$ and we
can verify that the typical fiber of $\mathcal{C}\ell_{C}(M,g)$ is
$\mathcal{C}\ell_{4,1}=\;\mathbb{C}\otimes\mathcal{C}\ell_{1,3}$, the Dirac
algebra. Now let $\{\Delta_{0},\Delta_{1},\Delta_{2},\Delta_{3},\Delta
_{4}\}\subset\sec\mathcal{C}\ell_{C}(M,g)$ be for all $x\in M$ an orthonormal
basis of $\mathcal{C}\ell_{4,1}$. We have,
\begin{align}
\Delta_{a}\Delta_{b}+\Delta_{b}\Delta_{a}  &  =2g_{ab}\;,\nonumber\\
&  g_{ab}=diag(+1,+1,+1,+1,-1)\;. \label{19}%
\end{align}

Let us identify $\gamma_{\mu}=\Delta_{\mu}\Delta_{4}$ and call $I=\Delta
_{0}\Delta_{1}\Delta_{2}\Delta_{3}\Delta_{4}$. Since $I^{2}=-1$ and $I$
commutes with all elements of $\mathcal{C}\ell_{4,1}$we identify $I$ with
$i=\sqrt{-1}$ and $\gamma_{\mu}$ with a fundamental set generating the local
Clifford algebra of $\mathcal{C}\ell(M,g)$. Then if $\mathcal{A}\in
\sec\mathcal{C}\ell_{C}(M,g)$ we have%

\begin{equation}
\mathcal{A}=\Phi_{s}+A_{C}^{\mu}\gamma_{\mu}+\frac{1}{2}B_{C}^{\mu\nu}%
\gamma_{\mu}\gamma_{\nu}+\frac{1}{3!}\tau_{C}^{\mu\nu\rho}\gamma_{\mu}%
\gamma_{\nu}\gamma_{\nu}+\Phi_{p}\gamma_{5}, \label{20}%
\end{equation}
where $\Phi_{s}$, $\Phi_{p}$, $A_{C}^{\mu}$, $B_{C}^{\mu\nu}$, $\tau_{C}%
^{\mu\nu\rho}\in\sec\;\mathbb{C}\otimes\bigwedge^{0}(M)\subset\sec
\mathcal{C}\ell_{C}(M,g)$, \emph{i.e.\/}, $\forall x\in M$, $\Phi_{s}(x)$,
$\Phi_{p}(x)$, $A_{C}^{\mu}(x)$, $B_{C}^{\mu\nu}(x)$, $\tau_{C}^{\mu\nu\rho
}(x)$ are complex numbers.

Now, it can be verified that
\begin{equation}
f=\frac{1}{2}(1+\gamma_{0})\frac{1}{2}(1+i\gamma_{1}\gamma_{2})\,;\quad
f^{2}=f\,, \label{21}%
\end{equation}
is a primitive idempotent field of $\mathcal{C}\ell_{C}(M,g)$. We can also
verify without difficulty that $if=\gamma_{2}\gamma_{1}f$.

Appropriate equivalence classes (see (\cite{16},\cite{17})) of $\mathcal{C}%
\ell_{C}(M,g)f$ are representatives of the standard Dirac spinor fields in
$\mathcal{C}\ell_{C}(M,g)$. We can easily show that the representation of
$\mathbf{\Psi}_{D}$ in $\mathcal{C}\ell_{C}(M,g)$ is given by
\begin{equation}
\mathbf{\Psi}_{D}=\psi f \label{22}%
\end{equation}
where $\psi$ is the Dirac-Hestenes spinor field given by Eq.(\ref{15}).

\subsection{Weyl Spinors and Parity Operator}

By definition, $\psi\in\sec\mathcal{C}\ell^{+}(M,g)$ is a \emph{representative
}of a Weyl spinor field (\cite{14},\cite{15}) if besides being a
representative of a Dirac-Hestenes spinor field it satisfies $\gamma_{5}%
\psi=\pm\psi\gamma$, where%
\begin{equation}
\gamma_{21}=\gamma_{2}\gamma_{1}. \label{22bis}%
\end{equation}
The positive (negative) \textquotedblleft eingestates\textquotedblright\ of
$\gamma_{5}$ will be denoted $\psi_{+}$ ($\psi_{-}$). For a general $\psi
\in\sec\mathcal{C}\ell^{+}(M,g)$ we can write
\begin{equation}
\psi_{\pm}=\frac{1}{2}\left[  \psi\mp\gamma_{5}\psi\gamma_{21}\right]  .
\label{22biss}%
\end{equation}

Then,
\begin{equation}
\psi=\psi_{+}+\psi_{-}. \label{22bisss}%
\end{equation}

The parity operator $\mathbf{P}$ in our formalism is represented in such a way
that for $\psi\in\sec\mathcal{C}\ell^{+}(M,g),$%
\begin{equation}
\mathbf{P}\psi=\mathbf{-}\gamma_{0}\psi\gamma_{0} \label{22BISSSS}%
\end{equation}

The following Dirac-Hestenes spinor fields are eingestates of the parity
operator with eingenvalues $\pm1$:
\begin{align}
\mathbf{P}\psi^{\uparrow}  &  =+\psi^{\uparrow},\text{ }\psi^{\uparrow}%
=\gamma_{0}\psi_{-}\gamma_{0}-\psi_{-},\nonumber\\
\mathbf{P}\psi^{\downarrow}  &  =-\psi^{\downarrow},\text{ }\psi^{\downarrow
}=\gamma_{0}\psi_{+}\gamma_{0}+\psi_{+} \label{22BF}%
\end{align}

\subsection{The spin-Dirac Operator}

Associated with the covariant derivative operator $D_{e_{a}}$ (see
Eq.(\ref{12n})) acting on sections of the Clifford bundle there is a
spin-covariant derivative operator $D_{e_{a}}^{s}$ acting on sections of a
right spin-Clifford bundle, such that its sections are Dirac-Hestenes spinor
fields. Hopefully it will be not necessary to present the details concerning
this concept here (see \cite{17}). Enough is to say that $D_{e_{a}}^{s}$ has a
representative on the Clifford bundle, called $D_{e_{a}}^{(s)}$, such that if
$\psi_{\Xi}$ is a \emph{representative} of a Dirac-Hestenes spinor field we
have
\begin{equation}
D_{e_{a}}^{(s)}\psi_{\Xi}=e_{a}(\psi_{\Xi})+\frac{1}{2}\mathbf{\omega}_{a}%
\psi_{\Xi}, \label{22n1}%
\end{equation}
where $\mathbf{\omega}_{a}$ has been defined by Eq.(\ref{12nn}). The
representative of the spin-Dirac operator acting on representatives of
Dirac-Hestenes spinor fields is the invariant first order operator given by,
\begin{equation}
{\mbox{\boldmath$\partial$}}^{(s)}=\gamma^{a}D_{e_{a}}^{(s)} \label{22nn}%
\end{equation}

>From the definition of spin-Dirac operator we see that if we restrict our
considerations to orthonormal coordinate bases $\{\gamma^{\mu}=dx^{\mu}\}$
where $\{x^{\mu}\}$ are global Lorentz coordinates then $\mathbf{\omega}_{\mu
}=0$ and the action of ${\mbox{\boldmath$\partial$}}^{(s)}$ on Dirac-Hestenes
spinor fields is the same as the action of ${\mbox{\boldmath$\partial$}}$ on
these fields.

\subsection{Maxwell and Dirac-Hestenes Equations}

With the mathematical tools presented above we have the following Maxwell
equation,
\begin{equation}
{\mbox{\boldmath$\partial$}}F=J_{e} \label{22nnm}%
\end{equation}
satisfied by an electromagnetic field $F\in\sec\bigwedge^{2}(M)\subset
\sec\mathcal{C}\!\ell(M,g)$, and generated by a current $J_{e}\in\sec
\bigwedge^{1}(M)\subset\sec\mathcal{C}\!\ell(M,g)$.

The Dirac-Hestenes equation in a spin frame $\Xi$ satisfied by a
Dirac-Hestenes spinor field $\psi\in\sec[\bigwedge^{0}(M)+\bigwedge
^{2}(M)+\bigwedge^{4}(M)]\subset\sec\mathcal{C}\!\ell(M,g)$ is
\begin{equation}
{\mbox{\boldmath$\partial$}}\psi\gamma^{2}\gamma^{1}-m\psi\gamma^{0}+\frac
{1}{2}\gamma^{a}\psi\mathbf{\omega}_{a}\gamma^{2}\gamma^{1}=0. \label{22nndh}%
\end{equation}

For what follows we restrict our considerations only for the case of
orthonormal coordinate basis, in which case the Dirac-Hestenes equation reads
\begin{equation}
{\mbox{\boldmath$\partial$}}\psi\gamma^{2}\gamma^{1}-m\psi\gamma^{0}=0
\label{22nndhb}%
\end{equation}

\section{Solution of $\psi\gamma_{21}\tilde{\psi}=F $}

We now want solve Eq.(\ref{1}) for $\psi$. Before proceeding we observe that
on Euclidian spacetime this equation has been solved using Clifford algebra
methods in \cite{29}. Also,on Minkowski spacetime a \emph{particular} solution
of an equivalent equation (written in terms of biquaternions) appear in
\cite{30}. We are going to show that contrary to the claims of \cite{21} a
general solution for $\psi$ has indeed eight degrees of freedom, although two
of them are \emph{arbitrary}, i.e., not fixed by $F$ alone. Once we give a
solution of Eq.(\ref{1}) for $\psi$, the reason for the indetermination of two
of the degrees of freedom will become clear. This involves the Fierz
identities, boomerangs (\cite{12},\cite{14},\cite{31}) and the general theorem
permitting the reconstruction of spinors from their bilinear covariants.

We start by observing that from Eq.(\ref{1}) and Eq.(\ref{16}) we can write
\begin{equation}
F=\rho e^{\beta\gamma_{5}}R\gamma_{21}\tilde{R} \label{23}%
\end{equation}

Then, defining $f=F/\rho e^{\beta\gamma_{5}}$ it follows that
\begin{align}
f  &  =R\gamma_{21}\tilde{R}\label{24}\\
f^{2}  &  =-1 \label{25}%
\end{align}

Now, since all objects in Eq.(\ref{23}) and Eq.(\ref{24}) are even we can take
advantage of the isomorphism $\mathcal{C}\!\ell_{3,0}\equiv\mathcal{C}%
\!\ell_{1,3}^{+}$ and making the calculations when convenient in the Pauli
algebra. To this end we first write:%

\begin{equation}
F=\frac{1}{2}F^{\mu\nu}\gamma_{\mu}\gamma_{\nu},\ F^{\mu\nu}=\left(
\begin{array}
[c]{cccc}%
0 & -E^{1} & -E^{2} & -E^{3}\\
E^{1} & 0 & -B^{3} & B^{2}\\
E^{2} & B^{3} & 0 & -B^{1}\\
E^{3} & -B^{2} & B^{1} & 0
\end{array}
\right)  , \label{26}%
\end{equation}
where $(E^{1},E^{2},E^{3})$ and $(B^{1},B^{2},B^{3})$ are respectively the
Cartesian components of the electric and magnetic fields.

We now write $F$ in $\mathcal{C}\ell^{+}(M,g)$, the even sub-algebra of
$\mathcal{C}\ell(M,g)$. The typical fiber of $\mathcal{C}\ell^{+}(M,g)$( which
is also a vector bundle) is isomorphic to the Pauli algebra. We put
\begin{equation}
\vec{\sigma}_{i}=\gamma_{i}\gamma_{0},\ \mathbf{i}=\vec{\sigma}_{1}\vec
{\sigma}_{2}\vec{\sigma}_{3}=\gamma_{0}\gamma_{1}\gamma_{2}\gamma_{3}%
=\gamma_{5}. \label{27}%
\end{equation}

Recall that $\mathbf{i}$ commutes with bivectors and since $\mathbf{i}^{2}=-1
$ it acts like the imaginary unit $i=\sqrt{-1}$ in $\mathcal{C}\ell^{+}(M,g)
$. From Eq.(\ref{26}) and Eq.(\ref{27}) (taking into account our previous
discussion) we can write
\begin{equation}
F=\vec{E}+\mathbf{i}\vec{B}, \label{28}%
\end{equation}
with $\vec{E}=E^{i}\vec{\sigma}_{i}$, $\vec{B}=B^{j}\vec{\sigma}_{j}$,
$i,j=1,2,3$. We can write an analogous equation for $f,$%
\begin{equation}
f=\vec{e}+\mathbf{i}\vec{b} \label{29}%
\end{equation}

Now, since $F^{2}\neq0$ and
\begin{align}
F^{2}  &  =F\lrcorner F+F\wedge F\nonumber\\
&  =(\vec{E}^{2}-\vec{B}^{2})+2\mathbf{i}(\vec{E}\cdot\vec{B}) \label{30}%
\end{align}
the above equations give (in the more general case where both $I_{1}=(\vec
{E}^{2}-\vec{B}^{2})\neq0$ and $I_{2}=(\vec{E}\cdot\vec{B})\neq0)$:
\begin{equation}
\rho=\frac{\sqrt{\vec{E}^{2}-\vec{B}^{2}}}{\cos[arctg2\beta]},\hspace
{0.25in}\beta=\frac{1}{2}\arctan\left(  \frac{2(\vec{E}\cdot\vec{B})}{\vec
{E}^{2}-\vec{B}^{2}}\right)  \label{31}%
\end{equation}

Also,
\begin{equation}
\vec{e}=\frac{1}{\rho}[(\vec{E}\cos\beta+\vec{B}\sin\beta)],\hspace
{0.25in}\vec{b}=\frac{1}{\rho}[(\vec{B}\cos\beta-\vec{E}\sin\beta)] \label{32}%
\end{equation}

\subsection{\bigskip A Particular Solution}

Now, we can verify that
\begin{align}
L  &  =\frac{\gamma_{21}+f}{\sqrt{2(1-\gamma_{5}\mathfrak{I})}}=\frac
{\vec{\sigma}_{3}-\mathbf{i}\vec{f}}{\mathbf{i}\sqrt{2(1-\mathbf{i}(\vec
{f}\cdot\vec{\sigma}_{3})}},\label{33}\\
\mathfrak{I}  &  =f^{03}-\gamma_{5}f^{12}\equiv\vec{f}\cdot\vec{\sigma}_{3}
\label{34}%
\end{align}
is a Lorentz transformation, i.e., $L\tilde{L}=\tilde{L}L=1$. Moreover, $L$ is
a particular solution of \ Eq.(\ref{24}). Indeed,
\begin{equation}
\frac{\gamma_{21}+f}{\sqrt{2(1-\gamma_{5}\mathfrak{I})}}\gamma_{21}%
\frac{\gamma_{12}-f}{\sqrt{2(1-\gamma_{5}\mathfrak{I})}}=\frac{f[2(1-\gamma
_{5}\mathfrak{I})]}{2(1-\gamma_{5}\mathfrak{I})}=f \label{35}%
\end{equation}

Of course, since $f^{2}=-1$, $\vec{e}^{2}=\vec{b}^{2}-1$and $\vec{e}\cdot
\vec{b}=0,$ there are only four real degrees of freedom in the Lorentz
transformation $L$. From this result in \cite{21} it is concluded that the
solution of the Eq.(\ref{1}) is the Dirac-Hestenes spinor field
\begin{equation}
\phi=\sqrt{\rho}e^{\gamma_{5}\beta}L, \label{36}%
\end{equation}
which has only \emph{six} degrees of freedom and thus is not equivalent to a
general Dirac-Hestenes spinor field (the spinor field that must appears in the
Dirac-Hestenes equation), which has \emph{eight} degrees of freedom. In this
way it is stated in \cite{21} that a the \emph{MDE} of first kind proposed in
(\cite{6},\cite{8}) cannot hold. Well, although it is \emph{true} that
Eq.(\ref{36}) is a solution of Eq.(\ref{1}) it is not a \emph{general}
solution, but only a \emph{particular} solution.

Before leaving this section we mention that there are many other Dirac-like
forms of the Maxwell equations published in the literature. All are trivially
related in a very simple way and in principle have nothing to do with the two
kinds of \emph{MDE} discussed in the present paper. See \cite{31}.

\subsection{The General Solution}

The general solution $R$ of Eq.(\ref{1}) is trivially found. It is
\begin{equation}
R=LS, \label{37}%
\end{equation}
where $L$ is the particular solution just found and $S$ is any member of the
\emph{stability group} of $\gamma_{21}$, i.e.,
\begin{equation}
S\gamma_{21}\tilde{S}=\gamma_{21},\text{ }S\tilde{S}=\tilde{S}S=1. \label{38}%
\end{equation}

\bigskip

\qquad It is trivial to find that we can parametrize the elements of the
stability group as
\begin{equation}
S=\exp(\gamma_{03}\nu)\exp(\gamma_{21}\varphi), \label{39}%
\end{equation}
with $0\leq\nu<\infty$ and $0\leq\varphi<\infty$. This shows that the most
\emph{general} Dirac-Hestenes spinor field that solves Eq.(\ref{1}) has indeed
eight degrees of freedom (as it must be the case, if the claims of (
\cite{6},\cite{8}) are to make sense), although two degrees of freedom are
arbitrary, i.e., they are like \emph{hidden variables}!

Now, the reason for the \emph{indetermination} of two degrees of freedom has
to do with a fundamental mathematical result: the fact that a spinor can only
be reconstruct through the knowledge of its bilinear covariants and the Fierz
identities. Explicitly, to reconstruct a Dirac-Hestenes spinor field $\psi$,
it is necessary to know also, besides the bilinear covariant given by Eq.(1),
the following bilinear covariants,
\begin{equation}
J=\psi\gamma_{0}\tilde{\psi}\text{ \textrm{and} }K=\psi\gamma_{3}\tilde{\psi}.
\label{40}%
\end{equation}

Now, $J,K$ and $F$ are related trough the so called Fierz identities,%

\begin{align}
J^{2}  &  =\sigma^{2}+\omega^{2}=-K^{2},\nonumber\\
J\cdot K  &  =0,\text{ }J\wedge K=-(\omega+\gamma_{5}\sigma)F,\nonumber\\
\sigma &  =\rho\cos\beta,\text{ }\omega=\rho\sin\beta. \label{41}%
\end{align}

In the most general case when both $\sigma,\omega$ are not $0$ we also have
the notable identities first found by Crawford \cite{31} (and which can be
derived almost trivially using the Clifford bundle formalism),
\begin{equation}%
\begin{array}
[c]{c}%
F\llcorner J=\omega K\\
(\gamma_{5}F)\llcorner J=\sigma K\\
F\cdot F=\langle F\tilde{F}\rangle_{0}=\sigma^{2}-\omega^{2}%
\end{array}
\hspace{0.25in}%
\begin{array}
[c]{c}%
F\llcorner K=\omega J\\
(\gamma_{5}F)\llcorner K=\sigma J\\
(\gamma_{5}F)\cdot F=2\sigma\omega
\end{array}
\label{42}%
\end{equation}

\begin{align}
JF  &  =(\omega+\gamma_{5}\sigma)K,\hspace{0.25in}KF=(\omega+\gamma_{5}%
\sigma)J\nonumber\\
F^{2}  &  =\omega^{2}-\sigma^{2}-2\gamma_{5}\sigma\omega,\hspace{0.25in}%
F^{-1}=KFK/(\omega^{2}+\sigma^{2})^{2} \label{43}%
\end{align}

Once we know $\omega$, $\sigma$, $J$, $K$ and $F$ we can recover the
Dirac-Hestenes spinor field as follows. First, introduce a \emph{boomerang
}(\cite{12},\cite{14},\cite{15}) $\mathfrak{B}\in\mathcal{C}\ell_{C}(M,g)$
given by
\begin{equation}
\mathfrak{B}=\sigma+J+iF-i\gamma_{5}K+\gamma_{5}\omega\label{44}%
\end{equation}

Then, we can construct $\mathbf{\Psi}=\mathfrak{B}f\in\mathcal{C}\ell
_{C}(M,g)f$ \ (with $f$ as in Eq.(\ref{21}))which has the following matrix
representation (once the standard representation of the Dirac gamma matrices
are used)
\begin{equation}
\mathbf{\hat{\Psi}}=\left(
\begin{array}
[c]{cccc}%
\psi_{1} & 0 & 0 & 0\\
\psi_{2} & 0 & 0 & 0\\
\psi_{3} & 0 & 0 & 0\\
\psi_{4} & 0 & 0 & 0
\end{array}
\right)  \label{45}%
\end{equation}

Now, it can be easily verified that $\mathbf{\Psi}=\mathfrak{B}f$ determines
the same bilinear covariants as the ones determined by $\psi$. Note however
that this spinor is not unique. In fact, $\mathfrak{B}$ determines a class of
elements $\mathfrak{B}\eta$ where $\eta$ is an arbitrary element of
$\mathcal{C}\ell_{C}(M,g)f$ which differs one from the other by a complex
phase factor (\cite{12}, \cite{14},\cite{15}).

Recalling that (a representative) of a Dirac-Hestenes spinor field determines
a unique element of $\Phi\in\mathcal{C}\ell_{C}(M)f$\ by $\Phi=\psi f$, then
it follows (from Eq.(\ref{45}) and Eq.(\ref{17}) that gives the matrix
representation of $\psi$) that we can trivially reconstruct a $\psi$ that
solves our problem.

\bigskip\ 

\section{The Generalized Maxwell Equation}

To comment on the basic error in \cite{21} concerning the Clifford bundle
formulation of the generalized Maxwell equation we recall the following.

The generalized Maxwell equation (\cite{9},\cite{31}) which describes the
electromagnetic field generated by charges and monopoles, can be written in
the Cartan bundle as%

\begin{equation}
dF=K_{m},\hspace{0.25in}dG=K_{e} \label{46}%
\end{equation}
where $F,G\in\bigwedge^{2}(M)$ and $K_{m},K_{e}\in\bigwedge\nolimits^{3}(M) $.

These equations are independent of \textit{any} metric structure defined on
the world manifold. When a metric is given and the Hodge dual operator $\star$
is introduced it is supposed that in vacuum we have $G=\star F$. In this case
putting $K_{e}=-\star J_{e}$ and $K_{m}=\star J_{m}$, with $J_{e},J_{m}\in
\sec\bigwedge^{1}(M)$, we can write the following equivalent set of equations
\begin{align}
dF  &  =-\star J_{m}\text{, }d\star F=-\star J_{e},\label{47a}\\
\delta(\star F)  &  =J_{m}\text{, }\delta F=-J_{e}\label{47b}\\
\delta(\star F)  &  =J_{m}\text{, }\delta F=-J_{e}\label{47c}\\
dF  &  =-\star J_{m}\text{, }\delta F=-J_{e}. \label{47d}%
\end{align}
Now, supposing that any $\sec\bigwedge^{j}(M)\subset\sec\mathcal{C}%
\!\ell(M,g)$ ($j=0,1,2,3,4$) and taking into account Eqs.(\ref{12}-\ref{13})
we get Eq.(\ref{2}) by summing the two equations in (\ref{47d}), i.e.,
\begin{equation}
(d-\delta\mathbf{)}F=J_{e}+K_{m}\text{ or}\hspace{0.25in}\text{ }%
(d-\delta\mathbf{)\star}F=-J_{m}+K_{e}, \label{48a}%
\end{equation}
or equivalently
\begin{equation}
{\mbox{\boldmath$\partial$}}F=J_{e}+\gamma_{5}J_{m}\text{ or}\hspace
{0.25in}\text{ }{\mbox{\boldmath$\partial$}}\mathbf{(-}\gamma_{5}%
F)=-J_{m}+\gamma_{5}J_{e}.\text{ } \label{48b}%
\end{equation}

\bigskip Now, writing \ with the conventions of section 2 ,
\begin{equation}
F=\frac{1}{2}F^{\mu\nu}\gamma_{\mu}\gamma_{\nu}\text{, }\star F=\frac{1}%
{2}(^{\star}F^{\mu\nu})\gamma_{\mu}\gamma_{\nu},\text{ } \label{49}%
\end{equation}
then generalized Maxwell equations in the form given by Eq.(\ref{47b}) can be
written in components ( in a Lorentz coordinate chart) as
\begin{equation}
\partial_{\mu}F^{\mu\nu}=J_{e}^{\mu}\text{, }\partial_{\mu}(^{\star}F^{\mu\nu
})=-J_{m}^{\mu} \label{50}%
\end{equation}
Now, assuming as in Eq.(\ref{1}) that $F=\psi\gamma_{21}\tilde{\psi}$ and
taking into account the relation between $\psi$ and the representation of the
\ standard Dirac spinor $\Psi_{D}$ given by Eq.(\ref{18}), we can write
Eq.(\ref{50}) as
\begin{align}
\partial_{\mu}\bar{\Psi}_{D}\left[  \hat{\gamma}_{\mu},\hat{\gamma}_{\nu
}\right]  \Psi_{D}  &  =2J_{e}^{\mu}\text{,\hspace{0.2in} }\partial_{\mu}%
\bar{\Psi}_{D}\hat{\gamma}_{5}\left[  \hat{\gamma}_{\mu},\hat{\gamma}_{\nu
}\right]  \Psi_{D}=-2J_{m}^{\mu},\nonumber\\
\hspace{0in}F^{\mu\nu}  &  =\frac{1}{2}\bar{\Psi}_{D}\left[  \hat{\gamma}%
_{\mu},\hat{\gamma}_{\nu}\right]  \Psi\text{, }(^{\star}F^{\mu\nu})=\frac
{1}{2}\bar{\Psi}_{D}\hat{\gamma}_{5}\left[  \hat{\gamma}_{\mu},\hat{\gamma
}_{\nu}\right]  \Psi_{D} \label{51}%
\end{align}

\bigskip

The reverse of the first of Eqs.(\ref{48b}) equation reads
\begin{equation}
\overset{\sim}{({\mbox{\boldmath$\partial$}}F)}=J_{e}-K_{m}. \label{52}%
\end{equation}

First summing, and then subtracting Eq.(\ref{2}) with Eq.(\ref{46}) we get the
following equations for $F=\psi\gamma_{21}\tilde{\psi}$,
\begin{equation}
{\mbox{\boldmath$\partial$}}\psi\gamma_{21}\tilde{\psi}+\overset{\sim
}{(\mathbf{\partial}\psi\gamma_{21}\tilde{\psi})}=2J_{e},\hspace
{0.25in}{\mbox{\boldmath$\partial$}}\psi\gamma_{21}\tilde{\psi}-\overset{\sim
}{(\mathbf{\partial}\psi\gamma_{21}\tilde{\psi})}=2K_{m} \label{53}%
\end{equation}
which is equivalent to Eq.(\ref{12}) in \cite{21} (where $\mathcal{G}$ is used
for the three form of monopolar current). There, it is observed that $J_{e}%
$\ is even under reversion and $K_{m}$ is odd. Then, it is claimed that
\textquotedblleft since reversion is a purely algebraic operation without any
particular physical meaning, the monopolar current $K_{m}$ is necessarily zero
if the Clifford formalism is assumed to provide a representation of Maxwell's
equation where the source currents $J_{e}$ and $K_{m}$ correspond to
fundamental physical fields.\textquotedblright\ It is also stated
that\ Eq.(\ref{51}) and Eq.(\ref{53}) imposes different constrains on the
monopolar currents $J_{e}$ and $K_{m}$.

It is clear that these arguments are fallacious. Indeed, it is obvious that if
any \emph{comparison} is to be made, it must be done between $J_{e}$ and
$J_{m}$ or between $K_{e}$ and $K_{m}$. In this case, it is obvious that both
pairs of currents have the same behavior under reversion. This kind of
confusion is widespread in the literature, mainly by people that works with
the generalized Maxwell equation(s) in component form (Eqs.(\ref{50})).

It seems that experimentally $J_{m}=0$ and the following question suggests
itself: is there any real physical field governed by a equation of the type of
the generalized Maxwell equation (Eq.(\ref{2})). The answer is \textit{yes}.

\section{The Generalized Hertz Potential Equation}

In what follows we accept that $J_{m}=0$ and take Maxwell equations for the
electromagnetic field $F\in\sec\bigwedge^{2}(M)\subset\sec\mathcal{C}%
\!\ell(M,g)$ and a current $J_{e}\in\sec\bigwedge^{1}(M)\subset\sec
\mathcal{C}\!\ell(M,g)$ as
\begin{equation}
{\mbox{\boldmath$\partial$}}F=J_{e}. \label{54}%
\end{equation}

Let $\Pi=\frac{1}{2}\Pi^{\mu\nu}\gamma_{\mu}\gamma_{\nu}=\vec{\Pi}%
_{e}+\mathbf{i}\vec{\Pi}_{m}\in\sec\bigwedge^{2}(M)\subset\sec\mathcal{C}%
\!\ell(M,g)$ be the so called \emph{Hertz potential} (\cite{33},\cite{34}). We
write
\begin{equation}
\left[  \Pi^{\mu\nu}\right]  =\left[
\begin{array}
[c]{cccc}%
0 & -\Pi_{e}^{1} & -\Pi_{e}^{2} & -\Pi_{e}^{3}\\
\Pi_{e}^{1} & 0 & -\Pi_{m}^{3} & \Pi_{m}^{2}\\
\Pi_{e}^{2} & \Pi_{m}^{3} & 0 & -\Pi_{m}^{1}\\
\Pi_{e}^{3} & -\Pi_{m}^{2} & \Pi_{m}^{1} & 0
\end{array}
\right]  . \label{55}%
\end{equation}
and define the \emph{electromagnetic potential }by
\begin{equation}
A=-\delta\Pi\in\sec\Lambda^{1}(T^{\star}M)\subset\sec\mathcal{C}\!\ell(M,g),
\label{56}%
\end{equation}

Since $\delta^{2}=0$ it is clear that $A$ satisfies the Lorenz gauge
condition, i.e.,
\begin{equation}
\delta A=0. \label{57}%
\end{equation}
Also, let
\begin{equation}
\gamma^{5}S=d\Pi\in\sec\bigwedge{}^{3}(M)\subset\sec\mathcal{C}\!\ell(M,g),
\label{58}%
\end{equation}
and call $S$, the \emph{Stratton potential}. It follows also that
\begin{equation}
d\left(  \gamma^{5}S\right)  =d^{2}\Pi=0. \label{59}%
\end{equation}
But $d(\gamma^{5}S)=\gamma^{5}\delta S$ from which we get, taking into account
Eq.(\ref{51}),
\begin{equation}
\delta S=0 \label{60}%
\end{equation}
We can put Eq.(\ref{56}) and Eq.(\ref{58}) in the form of a \emph{single}
generalized Maxwell like equation, i.e.,
\begin{equation}
{\mbox{\boldmath$\partial$}}\Pi=(d-\delta)\Pi=A+\gamma^{5}S=\mathcal{A}.
\label{61}%
\end{equation}
Eq.(\ref{61}) is the equation we were looking for. It is a legitimate physical
equation. We also have,
\begin{equation}
\square\Pi=(d-\delta)^{2}\Pi=dA+\gamma_{5}dS. \label{62}%
\end{equation}

Next, we define the electromagnetic field by
\begin{equation}
F={\mbox{\boldmath$\partial$}}\mathcal{A}=\square\Pi=dA+\gamma_{5}%
dS=F_{e}+\gamma_{5}F_{m}. \label{63}%
\end{equation}
We observe that,
\begin{equation}
\square\Pi=0\Rightarrow F_{e}=-\gamma_{5}F_{m}. \label{64}%
\end{equation}

Now, let us calculate $\mathbf{\partial}F$. We have,
\begin{equation}%
\begin{array}
[c]{rcl}%
{\mbox{\boldmath$\partial$}}F & = & (d-\delta)F\\[2ex]
& = & d^{2}A+d(\gamma^{5}dS)-\delta(dA)-\delta(\gamma^{5}dS).
\end{array}
\label{65}%
\end{equation}
The first and last terms in the second line of Eq.(\ref{62}) are obviously
null. Writing,
\begin{equation}
J_{e}=-\delta dA,\text{and }\gamma^{5}J_{_{m}}=-d(\gamma^{5}dS), \label{66}%
\end{equation}
we get Maxwell equation\emph{\ }
\begin{equation}
{\mbox{\boldmath$\partial$}}F=(d-\delta)F=J_{e}, \label{67}%
\end{equation}
if and only if the magnetic current $\gamma^{5}J_{m}=0$, i.e.,
\begin{equation}
\delta dS=0. \label{68}%
\end{equation}
a condition that we suppose to be satisfied in what follows. Then,
\begin{align}
\square A  &  =J_{e}=-\delta dA,\nonumber\\
\square S  &  =0. \label{69}%
\end{align}

Now, we define,
\begin{align}
F_{e}  &  =dA=\vec{E}_{e}+\mathbf{i}\vec{B}_{e},\label{70}\\[0.03in]
F_{m}  &  =dS=\vec{B}_{m}+\mathbf{i}\vec{E}_{m}. \label{71}%
\end{align}
and also
\begin{equation}
F=F_{e}+\gamma_{5}F_{m}=\vec{E}+\mathbf{i}\vec{B}=(\vec{E}_{e}-\vec{E}%
_{m})+\mathbf{i(}\vec{B}_{e}+\vec{B}_{m}). \label{72}%
\end{equation}
Then, we get
\begin{equation}
\square\vec{\Pi}_{e}=\vec{E},\quad\quad\square\vec{\Pi}_{m}=\vec{B}.
\label{73}%
\end{equation}

It is important to keep in mind that:
\begin{equation}
\square\Pi=0\Rightarrow\vec{E}=0\text{, and }\vec{B}=0. \label{74}%
\end{equation}
Nevertheless, despite this result we have,\medskip

\textbf{Hertz Theorem}
\begin{equation}
\square\Pi=0\text{ }\Longrightarrow\text{ }{\mbox{\boldmath$\partial$}}F_{e}=0
\label{75}%
\end{equation}

\textbf{Proof. } We have immediately from the above equations that
\begin{equation}
{\mbox{\boldmath$\partial$}}F_{e}=-\partial(\gamma_{5}F_{m})=-d(\gamma
_{5}dS)+\delta(\gamma_{5}dS)=\gamma_{5}d^{2}S-\gamma_{5}\delta
dS=0.\blacksquare\label{76}%
\end{equation}

We remark that Eq.(\ref{75}) has been called the Hertz theorem in
(\cite{33},\cite{35}) and it has been used there and \ also in (\cite{36}%
-\cite{42}) in order to find nontrivial \textit{superluminal }solutions of the
free Maxwell equation.

\section{Maxwell Dirac Equivalence of First Kind}

Let us consider a \emph{generalized} Maxwell equation%

\begin{equation}
{\mbox{\boldmath$\partial$}}F=\mathcal{J}\,, \label{6.13}%
\end{equation}
where $\mathbf{\partial}=\gamma^{\mu}\partial_{\mu}$ is the Dirac operator and
$\mathcal{J}$ is the electromagnetic current (an electric current $J_{e}$ plus
a magnetic monopole current $-\gamma_{5}J_{m}$, where $J_{e}$, $J_{m}\in
\sec\bigwedge^{1}M\subset{{\mathcal{C}\!\ell}}(M,g)$). We proved in section 2
that if $F^{2}\neq0$, then we can write%

\begin{equation}
F=\psi\gamma_{21}\tilde{\psi}\,, \label{6.14}%
\end{equation}
where $\psi\in\sec{{\mathcal{C}\!\ell}}^{+}(M,g)$ is a representative of a
Dirac-Hestenes field. If we use Eq.(\ref{6.14}) in Eq.(\ref{6.13}) we get%

\begin{equation}
{\mbox{\boldmath$\partial$}}(\psi\gamma_{21}\tilde{\psi})=\gamma^{\mu}%
\partial_{\mu}(\psi\gamma_{21}\tilde{\psi})=\gamma^{\mu}(\partial_{\mu}%
\psi\gamma_{21}\tilde{\psi}+\psi\gamma_{21}\partial_{\mu}\tilde{\psi
})=\mathcal{J}. \label{6.15}%
\end{equation}
from where it follows that
\begin{equation}
2\gamma^{\mu}\langle\partial_{\mu}\psi\gamma_{21}\tilde{\psi}\rangle
_{2}=\mathcal{J}, \label{6.16}%
\end{equation}
Consider the identity
\begin{equation}
\gamma^{\mu}\langle\partial_{\mu}\psi\gamma_{21}\tilde{\psi}\rangle
_{2}=\mathbf{\partial}\psi\gamma_{21}\tilde{\psi}-\gamma^{\mu}\langle
\partial_{\mu}\psi\gamma_{21}\tilde{\psi}\rangle_{0}-\gamma^{\mu}%
\langle\partial_{\mu}\psi\gamma_{21}\tilde{\psi}\rangle_{4}, \label{6.17}%
\end{equation}
and define moreover the vectors
\begin{equation}
j=\gamma^{\mu}\langle\partial_{\mu}\psi\gamma_{21}\tilde{\psi}\rangle_{0},
\label{6.18}%
\end{equation}%
\begin{equation}
g=\gamma^{\mu}\langle\partial_{\mu}\psi\gamma_{5}\gamma_{21}\tilde{\psi
}\rangle_{0}. \label{6.19}%
\end{equation}
Taking into account Eqs.(\ref{6.15})-(\ref{6.19}), we can rewrite
Eq.(\ref{6.15}) as
\begin{equation}
{\mbox{\boldmath$\partial$}}\psi\gamma_{21}\tilde{\psi}=\left[  \frac{1}%
{2}\mathcal{J}+\left(  j+\gamma_{5}g\right)  \right]  . \label{6.20}%
\end{equation}

Eq.(\ref{6.20}) is a \textit{spinorial representation }of Maxwell equation. In
the case where $\psi$ is non-singular (which corresponds to non-null
electromagnetic fields) we have
\begin{equation}
{\mbox{\boldmath$\partial$}}\psi\gamma_{21}=\frac{e^{\gamma_{5}\beta}}{\rho
}\left[  \frac{1}{2}\mathcal{J}+\left(  j+\gamma_{5}g\right)  \right]  \psi.
\label{6.21}%
\end{equation}

The Eq.(\ref{6.21}) representing Maxwell equation, written in that form, does
not appear to have any relationship with the Dirac-Hestenes equation
(Eq.(\ref{22nndhb})). However, we shall make some \emph{algebraic}
modifications on it in such a way as to put it in a form that suggest a very
interesting and \emph{intriguing relationship} between them, and consequently
a possible (?) connection between electromagnetism and quantum mechanics.

Since $\psi$ is supposed to be non-singular ($F\neq0$) we can use the
canonical decomposition of $\psi$ and write $\psi=\rho e^{\beta_{\gamma_{5}%
}/2}R$, with $\rho$, $\beta$ $\in\sec\bigwedge^{0}M\subset\sec\mathcal{C}%
\!\ell(M,g)$ and $R\in$ Spin$_{+}$(1,3), $\forall x\in M$. Then
\begin{equation}
\partial_{\mu}\psi=\frac{1}{2}(\partial_{\mu}\ln\rho+\gamma_{5}\partial_{\mu
}\beta+\Omega_{\mu})\psi, \label{6.22}%
\end{equation}
where we define the 2-form
\begin{equation}
\Omega_{\mu}=2(\partial_{\mu}R)\tilde{R}. \label{6.23}%
\end{equation}

Using this expression for $\partial_{\mu}\psi$ into the definitions of the
vectors $j$ and $g$ (Eqs.(\ref{6.18},\ref{6.19})) we obtain that
\begin{equation}
j=\gamma^{\mu}(\Omega_{\mu}\cdot S)\rho\cos\beta+\gamma_{\mu}[\Omega_{\mu
}\cdot(\gamma_{5}S)]\rho\sin\beta, \label{6.24}%
\end{equation}
\begin{equation}
g=[\Omega_{\mu}\cdot(\gamma_{5}S)]\rho\cos\beta-\gamma_{\mu}(\Omega_{\mu}\cdot
S)\rho\sin\beta, \label{6.25}%
\end{equation}
where we define the $spin$ 2-form $S$ by
\begin{equation}
S=\frac{1}{2}\psi\gamma_{21}\psi^{-1}=\frac{1}{2}R\gamma_{21}\tilde{R}.
\label{6.26}%
\end{equation}

We now define
\begin{equation}
J=\psi\gamma_{0}\tilde{\psi}=\rho v=\rho R\gamma^{0}R^{-1}, \label{6.27}%
\end{equation}
where $v$ is the \textit{velocity field} of the system. To continue, we define
the 2-form $\Omega=v^{\mu}\Omega_{\mu}$ and the scalars $\Lambda$ and $K$ by
\begin{equation}
\Lambda=\Omega\cdot S, \label{6.28}%
\end{equation}%
\begin{equation}
K=\Omega\cdot(\gamma_{5}S). \label{6.29}%
\end{equation}
\ Using these definition we have that
\begin{equation}
\Omega_{\mu}\cdot S=\Lambda v_{\mu}, \label{6.30}%
\end{equation}%
\begin{equation}
\Omega_{\mu}\cdot(\gamma_{5}S)=Kv_{\mu}, \label{6.31}%
\end{equation}
and for the vectors $j$ and $g$ can be written as
\begin{equation}
j=\Lambda v\rho\cos\beta+Kv\rho\sin\beta=\lambda\rho v, \label{6.32}%
\end{equation}%
\begin{equation}
g=Kv\rho\cos\beta-\Lambda v\rho\sin\beta=\kappa\rho v, \label{6.33}%
\end{equation}
where we defined
\begin{equation}
\lambda=\Lambda\cos\beta+K\sin\beta, \label{6.34}%
\end{equation}%
\begin{equation}
\kappa=K\cos\beta-\Lambda\sin\beta. \label{6.35}%
\end{equation}

The spinorial representation of Maxwell equation is written now as
\begin{equation}
{\mbox{\boldmath$\partial$}}\psi\gamma_{21}=\frac{e^{\gamma_{5}\beta}}{2\rho
}\mathcal{J}\psi+\lambda\psi\gamma_{0}+\gamma_{5}\kappa\psi\gamma_{0}.
\label{6.36}%
\end{equation}

Observe that there are (\cite{33}-\cite{41}) infinite families of non trivial
solutions of Maxwell equations such that $F^{2}\neq0$ (which correspond to
\emph{subluminal} and \emph{superluminal} solutions of Maxwell equation).
Then, it is licit to consider the case $\mathcal{J}=0$. We have,
\begin{equation}
{\mbox{\boldmath$\partial$}}\psi\gamma_{21}=\lambda\psi\gamma_{0}+\gamma
_{5}\kappa\psi\gamma_{0}, \label{6.37}%
\end{equation}
which is \textit{very} similar to the Dirac-Hestenes equation.

In order to go a step further into the relationship between those equations,
we remember that the electromagnetic field has \emph{six} degrees of freedom,
while a Dirac-Hestenes spinor field has \emph{eight} degrees of freedom and
that we proved in section 2 that two of these degrees of freedom are
\textit{hidden} variables. We are free therefore to impose two constraints on
$\psi$ if it is to represent an electromagnetic field. We choose these two
constraints as
\begin{equation}
{\mbox{\boldmath$\partial$}}\cdot j=0\text{ and }{\mbox{\boldmath$\partial$}}%
\cdot g=0. \label{6.38}%
\end{equation}

Using Eqs.(\ref{6.32},\ref{6.33}) these two constraints become
\begin{equation}
{\mbox{\boldmath$\partial$}}\cdot j=\rho\dot{\lambda}+\lambda
{\mbox{\boldmath$\partial$}}\cdot J=0, \label{6.39}%
\end{equation}%
\begin{equation}
{\mbox{\boldmath$\partial$}}\cdot g=\rho\dot{\kappa}%
+k{\mbox{\boldmath$\partial$}}\cdot J=0, \label{6.40}%
\end{equation}
where $J=\rho v$ and $\dot{\lambda}=(v\cdot{\mbox{\boldmath$\partial$}}%
)\lambda,$ $\dot{k}=(v\cdot{\mbox{\boldmath$\partial$}})k.$ These conditions
imply that
\begin{equation}
\kappa\lambda=\lambda\kappa\label{6.41}%
\end{equation}
which gives ($\lambda\neq0$):
\begin{equation}
\frac{\kappa}{\lambda}=const.=-\tan\beta_{0}, \label{6.42'}%
\end{equation}
or from Eqs.(\ref{6.34},\ref{6.35}):
\begin{equation}
\frac{K}{\Lambda}=\tan(\beta-\beta_{0}). \label{6.43}%
\end{equation}

Now we observe that $\beta$ is the angle of the duality rotation from $F$ to
$F^{\prime}=e^{\gamma_{5}\beta}F.$ If we perform another duality rotation by
$\beta_{0}$ we have $F\mapsto e^{\gamma_{5}(\beta+\beta_{0})}F,$ and for the
Takabayasi angle $\beta\mapsto\beta+\beta_{0}.$ If we work therefore with an
electromagnetic field duality rotated by an additional angle $\beta_{0}$, the
above relationship becomes
\begin{equation}
\frac{K}{\Lambda}=\tan\beta. \label{6.44}%
\end{equation}
This is, of course, just a way to say that we can choose the constant
$\beta_{0}$ in Eq.(\ref{6.42'}) to be zero. Now, this expression gives
\begin{equation}
\lambda=\Lambda\cos\beta+\Lambda\tan\beta\sin\beta=\frac{\Lambda}{\cos\beta},
\label{6.45}%
\end{equation}%
\begin{equation}
\kappa=\Lambda\tan\beta\cos\beta-\Lambda\sin\beta=0, \label{6.46}%
\end{equation}
and the spinorial representation of the Maxwell equation (Eq.(\ref{6.37}))
becomes
\begin{equation}
{\mbox{\boldmath$\partial$}}\psi\gamma_{21}-\lambda\psi\gamma_{0}=0
\label{6.47}%
\end{equation}
Note that $\lambda$ is such that
\begin{equation}
\rho\dot{\lambda}=-\lambda{\mbox{\boldmath$\partial$}}\cdot J. \label{6.48}%
\end{equation}

The current $J=\psi\gamma_{0}\tilde{\psi}$ is not conserved unless $\lambda$
is constant. If we suppose also that
\begin{equation}
{\mbox{\boldmath$\partial$}}\cdot J=0 \label{6.49}%
\end{equation}
we must have
\[
\lambda=\text{const.}%
\]

Now, throughout these calculations we have assumed $\hbar=c=1$. We observe
that in Eq.(\ref{6.47}) $\lambda$ has the units of (length)$^{-1}$, and if we
introduce the constants $\hbar$ and $c$ we have to introduce another constant
with unit of mass. If we denote this constant by $m$ such that
\begin{equation}
\lambda=\frac{mc}{\hbar}, \label{6.50}%
\end{equation}
then Eq.(\ref{6.47}) assumes a form which is identical to Dirac-Hestenes
equation:
\begin{equation}
{\mbox{\boldmath$\partial$}}\psi\gamma_{21}-\frac{mc}{\hbar}\psi\gamma_{0}=0.
\label{6.51}%
\end{equation}

It is true that we didn't prove that Eq.(\ref{6.51}) is really Dirac equation
since the constant $m$ has to be identified in this case with the electron's
mass, and we do not have \textit{any} good physical argument to make that
identification, until now. In resume, Eq.(\ref{6.51}) has been obtained from
Maxwell equation by imposing some gauge conditions allowed by the hidden
parameters in the solution of Eq.(\ref{1}) for $\psi$ in terms of $F$. In view
of that, it seems more appropriate instead of using the term
\textit{mathematical }Maxwell-Dirac equivalence of first kind to talk about a
correspondence between that equations \textit{under }which the two extra
degrees of freedom of the Dirac-Hestenes spinor field are treated as hidden variables.

To end this section we observe that it is to earlier to know if the above
results are of some physical value or only a mathematical curiosity. Let us wait...

\section{Maxwell-Dirac Equivalence of Second Kind}

We now look for a Hertz potential field $\Pi\in\sec\bigwedge\nolimits^{2}(M)$
satisfying the following (\emph{non linear}) equation
\begin{equation}
{\mbox{\boldmath$\partial$}}\Pi=({\mbox{\boldmath$\partial$}}\mathfrak{G}%
+m\mathfrak{P}\gamma_{3}+m\langle\Pi\gamma_{012}\rangle_{1})+\gamma
_{5}({\mbox{\boldmath$\partial$}}\mathfrak{P}+m\mathfrak{G}\gamma_{3}%
-\gamma_{5}\langle m\Pi\gamma_{012}\rangle_{3}) \label{7.1}%
\end{equation}
where $\mathfrak{G,P}\in\sec\bigwedge\nolimits^{0}(M)$, and $m$ is a constant.
According to section 5 the electromagnetic and Stratton potentials are
\begin{equation}
A={\mbox{\boldmath$\partial$}}\mathfrak{G}+m\mathfrak{P}\gamma_{3}+m\langle
\Pi\gamma_{012}\rangle_{1,} \label{7.2}%
\end{equation}%
\begin{equation}
\gamma_{5}S=\gamma_{5}({\mbox{\boldmath$\partial$}}\mathfrak{P}+m\mathfrak{G}%
\gamma_{3}-\gamma_{5}\langle m\Pi\gamma_{012}\rangle_{3}), \label{7.3}%
\end{equation}
and must satisfy the following subsidiary conditions,
\begin{equation}
\square({\mbox{\boldmath$\partial$}}\mathfrak{G}+m\mathfrak{P}\gamma
_{3}+m\langle\Pi\gamma_{012}\rangle_{1})=J_{e} \label{7.4}%
\end{equation}%
\begin{equation}
\square(\gamma_{5}({\mbox{\boldmath$\partial$}}\mathfrak{P}+m\mathfrak{G}%
\gamma_{3}-\gamma_{5}\langle m\Pi\gamma_{012}\rangle_{3}))=0, \label{7.5}%
\end{equation}%
\begin{equation}
\square\mathfrak{G}+m{\mbox{\boldmath$\partial$}}\cdot\langle\Pi\gamma
_{012}\rangle_{1}=0, \label{7.6}%
\end{equation}%
\begin{equation}
\square\mathfrak{P}-m{\mbox{\boldmath$\partial$}}\cdot(\gamma_{5}\langle
\Pi\gamma_{012}\rangle_{3})=0. \label{7.7}%
\end{equation}

Now, in the Clifford bundle formalism, as we already explained above, the
following sum is a legitimate operation
\begin{equation}
\psi=-\mathfrak{G}+\Pi+\gamma_{5}\mathfrak{P} \label{7.8}%
\end{equation}
and according to the results of section 2 defines $\psi$ as a (representative)
of \ some Dirac-Hestenes spinor field. Now, we can verify that $\psi$
satisfies the equation
\begin{equation}
{\mbox{\boldmath$\partial$}}\psi\gamma_{21}-m\psi\gamma_{0}=0 \label{7.9}%
\end{equation}
which is as we already know a \emph{representative} of the standard Dirac
equation (for a free electron) in the Clifford bundle, which is a
Dirac-Hestenes equation (Eq.(\ref{22nndhb})), written in an orthonormal
coordinate \ spin frame.

The above developments suggest (consistently with the spirit of the
generalized Hertz potential theory developed in section 5) the following
interpretation. The Hertz potential field $\Pi$ generates the real
electromagnetic field of the electron (The question of the physical dimensions
of the Dirac-Hestenes and Maxwell fields is discussed in \cite{8}.) Moreover,
the above developments suggest that the electron is \textquotedblleft
composed\textquotedblright\ of two \textquotedblleft
fundamental\textquotedblright\ currents, one of \emph{electric} type and the
\emph{other} of magnetic type circulating at the ultra microscopic level,
which generate the observed electric charge and magnetic moment of the
electron. Then, it may be the case, as speculated by Maris \cite{27}, that the
electromagnetic field of the electron can be spliced into two parts, each
corresponding to a new kind of subelectron type particle, the \emph{electrino}%
. Of course, the above developments leaves open the possibility to generate
electrinos of fractional charges. We still study more properties of the above
system in another paper.

\section{Seiberg-Witten Equations}

As it is well known, the original Seiberg-Witten (monopole) equations have
been written in euclidean \textquotedblleft spacetime\textquotedblright\ and
for the \textit{self dual }part of the field $F$. However, on Minkowski
spacetime, of course, there are \textit{no} self dual electromagnetic fields.
Indeed, Eq.(\ref{11b}) implies that the unique solution (on Minkowski
spacetime) of the equation $\star F=F$ is $F=0$. This is the main reason for
the difficulties in interpreting that equations in this case, and indeed in
\cite{28} it was attempted an interpretation of that equations only for the
case of euclidean manifolds. Here we want to derive and to give a possible
interpretation to that equations based on a reasonable assumption.

Now, the \textit{analogous} of Seiberg-Witten monopole equations \textit{read}
in the Clifford bundle formalism and on Minkowski spacetime as
\begin{equation}
\left\{
\begin{array}
[c]{l}%
{\mbox{\boldmath$\partial$}}\psi\gamma_{21}-A\psi=0\\
F=\frac{1}{2}\psi\gamma_{21}\tilde{\psi}\\
F=dA
\end{array}
\right.  \label{8.1}%
\end{equation}
where $\psi\in\sec\mathcal{C}\ell^{+}(M,g)$ is a Dirac-Hestenes spinor field,
$A\in\sec\bigwedge\nolimits^{1}(M)\subset\sec\mathcal{C}\ell(M,g)$ is an
electromagnetic vector potential and $F\in\sec\bigwedge\nolimits^{2}%
(M)\subset\sec\mathcal{C}\ell(M,g)$ is an electromagnetic field.

Our intention in this section is:

(a) To use the Maxwell Dirac-Equivalence of the first kind (proved in section
7) and an additional hypothesis to be discussed below to derive the
Seiberg-Witten equations on Minkowski spacetime.

(b) to give a (possible) physical interpretation for that equations.

\subsection{Derivation of Seiberg-Witten Equations}

\textbf{Step 1.} We assume that the electromagnetic field $F$ appearing in the
second of the Seiberg-Witten equations satisfy the free Maxwell equation,
i.e., $\partial F=0.$

\textbf{Step 2. }We use the Maxwell-Dirac equivalence of the first kind proved
in section 6 to obtain Eq.(\ref{6.47}),
\begin{equation}
{\mbox{\boldmath$\partial$}}\psi\gamma_{21}-\lambda\psi\gamma_{0}=0
\label{8.2}%
\end{equation}

\textbf{Step 3.} We introduce the \textit{ansatz}
\begin{equation}
A=\lambda\psi\gamma_{0}\psi^{-1}. \label{8.3}%
\end{equation}

This means that the electromagnetic potential (in our geometrical units) is
identified with a multiply of the velocity field defined through
Eq.(\ref{6.27}). Under this condition Eq.(\ref{8.2}) becomes
\begin{equation}
{\mbox{\boldmath$\partial$}}\psi\gamma_{21}-A\psi=0, \label{8.4}%
\end{equation}
which is the first Seiberg-Witten equation!

\subsection{A Possible Interpretation of the Seiberg-Witten Equations}

Well, it is time to find an interpretation for Eq.(\ref{8.4}). In order to do
that we recall from section 2.5 that if $\psi_{\pm}$ are Weyl spinor fields
(as defined through Eq.(\ref{22bis}), then $\psi_{\pm}$ satisfy a Weyl
equation, i.e.,
\begin{equation}
{\mbox{\boldmath$\partial$}}\psi_{\pm}=0. \label{8.5}%
\end{equation}

Consider now, the equation for $\psi_{+}$ coupled with an electromagnetic
field $A=gB\in\sec\bigwedge\nolimits^{1}(M)\subset\sec\mathcal{C}\ell(M,g)$,
i.e.,
\begin{equation}
{\mbox{\boldmath$\partial$}}\psi_{+}\gamma_{21}+gB\psi_{+}=0. \label{8.6}%
\end{equation}

This equation is invariant under the gauge transformations
\begin{equation}
\psi_{+}\mapsto\psi_{+}e^{g\gamma_{5}\theta};B\mapsto B+\partial\theta.
\label{8.7}%
\end{equation}

Also, the equation for $\psi_{-}$ coupled with an electromagnetic field
$gB\in\sec\bigwedge\nolimits^{1}(M)$ is
\begin{equation}
{\mbox{\boldmath$\partial$}}\psi\gamma_{21}+gB\psi_{-}=0. \label{8.8}%
\end{equation}
which is invariant under the gauge transformations
\begin{equation}
\psi_{-}\mapsto\psi_{-}e^{g\gamma_{5}\theta};B\mapsto B-\partial\theta.
\label{8.9}%
\end{equation}
showing clearly that the fields $\psi_{+}$ and $\psi_{-}$ carry
\textit{opposite} `charges'. Consider now the Dirac-Hestenes spinor fields
$\psi^{\uparrow},\psi^{\downarrow}$ given by Eq.(\ref{22BF}) which are
eigenvectors of the \emph{parity} operator and look for solutions of
Eq.(\ref{8.4}) such that $\psi=\psi^{\uparrow}.$ We have,
\begin{equation}
{\mbox{\boldmath$\partial$}}\psi^{\uparrow}\gamma_{21}+gB\psi^{\uparrow}=0
\label{8.10}%
\end{equation}
which separates in two equations,
\begin{equation}
{\mbox{\boldmath$\partial$}}\psi_{+}^{\uparrow}+g\gamma_{5}B\psi_{+}%
^{\uparrow}=0;\text{ }{\mbox{\boldmath$\partial$}}\psi_{-}^{\uparrow}%
-g\gamma_{5}B\psi_{-}^{\uparrow}=0. \label{8.11}%
\end{equation}

These results show that when a Dirac-Hestenes spinor field associated with the
first of the Seiberg-Witten equations is in an eigenstate of the parity
operator, that spinor field describes a \textit{pair} of particles with
opposite `charges'. We interpret these particles (following Lochack \cite{42},
that suggested that an equation equivalent to Eq.(\ref{8.11}) describe
massless monopoles of opposite `charges') as being \emph{massless} `monopoles'
in \emph{auto-interaction}. Observe that our proposed interaction is also
consistent with the third of Seiberg-Witten equations, for $F=dA$ implies a
\emph{null} magnetic current.

It is now well known that Seiberg-Witten equations have non trivial solutions
on Minkowski manifolds (see \cite{25}). From the above results, in particular,
taking into account the inversion formula (Eq.(\ref{35})) it seems to be
possible to find whole family of solutions for the Seiberg-Witten equations,
which has been here derived from a Maxwell-Dirac equivalence of first kind
(proved in section 6) with the additional hypothesis that electromagnetic
potential $A$ is parallel to the velocity field $v$ (Eq.(\ref{8.3})) of the
system described by Eq.(\ref{6.27}). We conclude that a consistent set of
Seiberg-Witten equations on Minkowski spacetime must be
\begin{equation}
\left\{
\begin{array}
[c]{l}%
{\mbox{\boldmath$\partial$}}\psi\gamma_{21}-A\psi=0\\
F=\frac{1}{2}\psi\gamma_{21}\tilde{\psi}\\
F=dA\\
A=\lambda\psi\gamma_{0}\psi^{-1}%
\end{array}
\right.  \label{8.12}%
\end{equation}

\section{Conclusions}

In this paper we exhibit two different kinds of possible Maxwell-Dirac
equivalences (\emph{MDE}). Although many will find the ideas presented above
speculative from the physical point of view, we hope that they may become
important, at least from a mathematical point of view. Indeed, not to long
ago, researching solutions of the free Maxwell equation ($\partial F=0$)
satisfying the constraint $F^{2}\neq0$ (a necessary condition for derivation
of a \emph{MDE} of the first kind) conduced to the \textit{discovery} of
families of \emph{superluminal} solutions of Maxwell equations and also of all
the main linear relativistic equations of theoretical Physics (\cite{34}%
,\cite{42}). The study of the \emph{MDE} of the second kind reveal an
unsuspected interpretation of the Dirac equation, namely that the electron
seems to be a composed system build up from the self interaction of two
currents of `electrical' and `magnetic' types. Of course, it is to earlier to
say if this discovery has any physical significance. We showed also, that by
using the \emph{MDE} of the first kind together with a reasonable hypothesis
we can shed light on the meaning of Seiberg-Witten monopole equations on
Minkowski spacetime. We hope that the results just found may be an indication
that Seiberg-Witten equations (which are a fundamental key in the study of the
topology of four manifolds equipped with an \textit{euclidean} metric tensor),
may play an important role in Physics, whose arena where phenomena occur is a
\textit{Lorentzian} manifold.\medskip

\textbf{acknowledgements}: The author is grateful to Drs. V. V. Fern\'{a}ndez,
A. Gsponer, R. A. Mosna, A. M. Moya, I. R. Porteous and J. Vaz Jr. for useful
discussions. He is grateful also to CNPq for a senior research fellowship
(contract 201560/82-8) and to the Department of Mathematical Sciences,
University of Liverpool for a Visiting Professor position during the academic
year 2001/2002 and hospitality.

\end{document}